\documentstyle[aps,epsfig]{revtex}
\newcommand \la{\raisebox{-.5ex}{$\stackrel{<}{\sim}$}}
\begin{document}
%%%%%%%%%%%%%%%%%%%%%%%%%%%%%%%%%%%%%%%%%%%%%%%%%%%%%%%%%%%%%%%%%%%%%%%%%%%%%%%
\begin{titlepage}
\pagestyle{empty}
\baselineskip=18pt
\rightline{NBI--97--32}
\rightline{July, 1997}
\baselineskip=15pt
\vskip .2in
\begin{center}
{\large{\bf The Phase in Three--Pion Correlations}}
\end{center}
\vskip .2truecm
\begin{center}
Henning Heiselberg 

{\it NORDITA, Blegdamsvej 17, DK-2100 Copenhagen \O., Denmark}

and Axel P.~Vischer

{\it Niels Bohr Institute, DK-2100, Copenhagen \O, Denmark.}

\end{center}

\vskip 0.1in
\centerline{ {\bf Abstract} }
\baselineskip=15pt
\vskip 0.5truecm

We discuss the complex phase generated in three pion correlation
functions. The lowest order contribution to the phase is of order
$q^2 R/K$, where $q$ is a typical relative momentum, $K$ is a
typical center of mass momentum and $R$ is a typical radius parameter.
This contribution is of purely kinematic origin. At next order we find
a generic contribution of order $(qR)^3$ which is a result of odd
modifications to the source emission function. We argue, that the
scale for typical HBT correlations in ultrarelativistic heavy ion
collisions is $q/K \ll qR \sim 1$, so that the third order correction
actually dominates the phase in the experimentally relevant momentum
range. We study in detail such contributions which arise from source
asymmetries generated by flow, the source geometry and resonance
decays.
\end{titlepage} 
%%%%%%%%%%%%%%%%%%%%%%%%%%%%%%%%%%%%%%%%%%%%%%%%%%%%%%%%%%%%%%%%%%%%%%%%%

\baselineskip=15pt
\textheight8.9in\topmargin-0.0in\oddsidemargin-.0in

\section{Introduction}

Recently the first study of three pion correlations ($\pi^+
\pi^+ \pi^+$) in heavy ion collisions was reported by
NA44~\cite{janus}. The quality of the data was rather limited
due to statistics, but we can expect a steady
increase in statistics in forthcoming experiments. This should allow us
to explore large parts of the phase
space of the three pion correlation function in the foreseeable
future. 

The theoretical understanding of the three--pion correlation function
is still rather limited. In an earlier paper~\cite{hv1} we constructed
the three pion correlation function under the assumption that the
source was incoherent and only interfered through Bose-Einstein
correlations. Thus the three-particle correlation function could be
derived from the two-body correlation function 
except for a complex phase, which will be
studied here. Any deviation found experimentally from this prediction
would have signalled new physics. We found satisfactory agreement with
the available data, but due to the poor statistics could not make any
conclusive statements.

Heinz and Zhang~\cite{HZ} explored the structure of the complex phase
using an expansion in terms of the relative momentum of the two
emitted particles. They found
the lowest order contribution to be in the order $q^2 R/K$. 
Here, the relative momentum is given by ${\bf q}=({\bf k}_1-{\bf k}_2)$
the average momentum by ${\bf K}=({\bf k}_1+{\bf k}_2)/2$ and ${\bf
k}_i$ is the single particle momentum of particle $i$. This
contribution is of purely kinematic origin. Typical heavy
ion sources in nuclear collisions are of size $R\sim5$ fm, so that
interference occurs predominantly when 
$q\raisebox{-.5ex}{$\stackrel{<}{\sim}$}\hbar/R\sim 40$
MeV/c. Since typical particle momenta are $k_i\simeq K\sim 300$ MeV,
we find that $q/K \ll 1$ and the lowest order contribution to the
phase is very small. 

In this paper we will extend the two calculations mentioned above and 
explicitly investigate the modifications due to generic three pion
correlations. In section 2 we review the general derivation of the
phase in 3 particle
correlations in terms of an expansion up to cubic terms in the
relative momenta $q$. In section 3 we estimate the size of 
the phase for a number of physical effects known to be present in
relativistic heavy ion collisions and which produce asymmetric sources
leading to a complex phase in the 3 particle
correlation function. These are flow, resonance decay and asymmetric 
source geometry. 
We find some contributions to be sizable, but the overall
effect off such a phase on the measurable correlation function turns
out to be rather small and any detection will require exceedingly good
resolution in the regime for which $qR\sim\hbar$.
Finally, in the conclusions, we summarise our results and 
discuss some experimental consequences.

\section{General Formalism}

The three pion correlation function for incoherent sources
is given by~\cite{hv1,cramer,biya,weiner}
\begin{eqnarray} 
C_3(k_1,k_2,k_3)
             &=&1+F_{12}\;F_{21}+F_{23}\;F_{32}+F_{31}\;F_{13}
                  +F_{12}\;F_{23}\;F_{31}
		+F_{21}\;F_{32}\;F_{13}\;\nonumber\\
&=&1+|F_{12}|^2+|F_{23}|^2+|F_{31}|^2
                  +2 \;Re\left[F_{12}\;F_{23}\;F_{31}\right]\;,
\label{3c}
\end{eqnarray}
where $F_{ij}$ is the Fourier transform of the source emission
function $S(x,K)$~\cite{def} 
\begin{eqnarray}
F_{ij}\equiv F(q_{ij},K_{ij})=\frac{\int d^4 x\;S(x,K_{ij})\;\exp(i q_{ij} x)}
	{\sqrt{\int d^4 x\;S(x, k_i)\int d^4 x\;S(x, k_j)}}=F_{ji}^{\star}\;.
                         \label{fs}
\end{eqnarray}
We used here the relative momentum $q_{ij} = k_i-k_j$ and the center
of momentum variable $K_{ij}~=~(k_i~+~k_j)/2$. The emission function
$S(x,k)$ is the probability of emission for a pion from space--time 
point $x$ with
momentum $k$. It is related to the experimentally measured pion single 
particle spectrum 
\begin{eqnarray}
E_k\;\frac{dN}{d^3 k} = \int d^4 x\;S(x,k)\;,
\label{sps}
\end{eqnarray}
where $E_k = k_0 = \sqrt{{\bf k}^2+m^2}$ is the on mass--shell energy. 

The relative momenta of three particles satisfy the relation
\begin{equation}
q_{12}+q_{23}+q_{31}=0\;,
\label{triangle}
\end{equation}
i.e., they span a triangle. This automatically assures translational
invariance of the 3-body correlation function. Any translation in
space-time by a distance $x_0$ will lead to an extra phase factor 
$exp(iq_{ij}\cdot x_0)$ in $F_{ij}$ but Eq. (\ref{triangle}) insures
that the triple product of the three phase factors cancel in the
3-body correlation function (\ref{3c}).

The source emission function in the numerator of Eq. 
(\ref{fs}) is not evaluated for a
momentum $k_i$, but rather for the center of mass momentum
$K_{ij}$. Since all particles are detected on shell, we will have
to evaluate the center of mass momenta in (\ref{fs}) slightly off
shell
\begin{eqnarray}
K_{ij}^0 = E_K \;\left(1+\frac{{\bf q}^2}{8\,E_K^2}+{\cal O}
\left(\frac{{\bf q}^4}{\,E_K^4}\right)\right)\;.
\label{massshell}
\end{eqnarray}
This on--shell constraint for the two detected particles pushes the
source emission function slightly off--shell by an amount
\begin{eqnarray}
S(x,K_{ij}^0,{\bf K}_{ij}) = S(x,E_K,{\bf K}_{ij})\;
+\frac{{\bf q}_{ij}^2}{8\,E_K}\;\frac{\partial S}{\partial
E_K}(x,E_K,{\bf K}_{ij})
+{\cal O}({\bf q}_{ij}^4)\;.
\label{massexp}
\end{eqnarray}
This correction is of order $(q/K)^2$ and is also present in the
2--particle correlation function. Since the off--shell structure
of the source emission function is not accessible to us we will
neglect this contribution, like in the 2--particle case, 
with the remark that a possible signal
could be due to this correction.

With this in mind, we can rewrite Eq. (\ref{3c}) using on shell 
variables $F_{ij}=F({\bf q}_{ij},{\bf K}_{ij})$ 
and explicitly depict the phase $\phi_{ij}=\phi({\bf q}_{ij},{\bf K}_{ij})$
\begin{eqnarray}
C_3({\bf k}_1,{\bf k}_2,{\bf k}_3)=1
+|F_{12}|^2+|F_{23}|^2+|F_{31}|^2 +2\;|F_{12}|\;|F_{23}|\;|F_{31}|
\;\cos{(\phi_{12}+\phi_{23}+\phi_{31})}  \;.
\label{c3p}
\end{eqnarray}
All information about possible imaginary contributions to $F$ are now 
contained in the cosine of the 3 phases, $\phi_{ij}$, defined as
\begin{eqnarray}
\tan{\phi_{ij}} 
= \frac{Im\left[F_{ij}\right]}{Re\left[F_{ij}\right]}\;.
\label{phi}
\end{eqnarray}
The phases arise from the triple
product of Fourier transforms in Eq. (\ref{3c}). For the
following discussion it is important to realize, that the three
particle correlations as given above are defined over a 9 dimensional 
momentum space. This space can be either described by the 3 vectors
${\bf k}_1,{\bf k}_2,{\bf k}_3$, or preferably by the center of
momentum of the three emitted particles ${\bf K} = ({\bf k}_1 + 
{\bf k}_2 + {\bf k}_3)/3$ and two relative momenta, like ${\bf q}_{12}$
and ${\bf q}_{23}$. The kinematic transformation into center of
momentum and relative momenta will actually fix the lowest order
contribution in our expansion below.

To determine the phases we have to extract both the real and
imaginary pieces of the Fourier transforms of the source emission 
function. The
source emission function itself is real. The only source for an imaginary
contribution is thus the Fourier transform via the exponential of
the relative momentum. 
We introduce the symmetric and antisymmetric part, $S_s$ and $S_a$,
of the source emission function by
\begin{eqnarray}
S_s\left(x-\langle x \rangle,{\bf K}\right) &=& 
\frac{1}{2} \left[S\left(x-\langle x \rangle,{\bf K}\right)
+S\left(-(x-\langle x \rangle),{\bf K}\right)\right]\nonumber\\
S_a\left(x-\langle x \rangle,{\bf K}\right) &=& 
\frac{1}{2} \left[S\left(x-\langle x \rangle,{\bf K}\right)
-S\left(-(x-\langle x \rangle),{\bf K}\right)\right]\;.
\end{eqnarray}
The average of an operator
$\xi$ is defined as~\cite{Heinz}
\begin{eqnarray}
\langle \hat{\xi} \rangle = \frac{\int d^4 x\;\hat{\xi}\;S(x,K)} 
{\int d^4 x\;S(x,K)}\;.
\end{eqnarray}
The space--time integral over the asymmetric
part is identical zero, while the space--time integral over the
symmetric part provides the normalisation of the source emission
function. 

The phase $\phi_{ij}$ in Eq. (\ref{phi}) is now
\begin{eqnarray}
\tan{\phi_{ij}} 
= \frac{1}{i}\; \frac{\int d^4 x\;S_a(x,E_K,{\bf K}_{ij})
\;\exp(i q_{ij} x)}{\int d^4 x\;S_s(x,E_K,{\bf K}_{ij})\;\exp(i q_{ij} x)}\;,
\label{phiprime}
\end{eqnarray}
which clearly demonstrates that
the odd space--time moments of the source emission 
function $S$ generates the phase. If $S$ is even, then the
Fourier transform in (\ref{fs}) will be real and the phases $\phi_{ij}$
vanish. 

If we expand the exponential in Eq. (\ref{phiprime}) for small values of
$q_{ij}$
\begin{equation}
\exp{(i q_{ij} x)} = 1+i q_{ij} x-\frac{1}{2} \;(q_{ij}
x)^2-\frac{i}{6}\; (q_{ij} x)^3+{\cal O}(q_{ij}^4)\;,
\end{equation}
we obtain for the phase
\begin{eqnarray}
\tan{\phi_{ij}}=\langle (q_{ij} x) \rangle
-\frac{1}{6}\; \langle (q_{ij} (x-\langle x \rangle))^3 \rangle
\rangle 
+{\cal O}(q_{ij}^4)\;.
\label{expphi}
\end{eqnarray}
The sum of the 3 phases is then
\begin{eqnarray}
\phi_{12}+\phi_{13}+\phi_{23} &=&
 \frac{1}{2}\;q_{12}^{\mu} q_{23}^{\nu} \left[
\frac{\partial \langle x_{\mu} \rangle}{\partial K^{\nu}} 
-\frac{\partial \langle x_{\nu} \rangle}{\partial K^{\mu}} \right]\nonumber\\
&-&\frac{1}{24}\;\left[q_{12}^{\mu}  q_{12}^{\nu} q_{23}^{\lambda}
+q_{23}^{\mu}  q_{23}^{\nu} q_{12}^{\lambda}\right]
\left[\frac{\partial^2 \langle x_{\mu} \rangle}
{\partial K^{\nu}\partial K^{\lambda}}
+\frac{\partial^2 \langle x_{\nu} \rangle}
{\partial K^{\lambda}\partial K^{\mu}}
+\frac{\partial^2 \langle x_{\lambda} \rangle}
{\partial K^{\mu}\partial K^{\nu}} \right] \nonumber \\
&-&\frac{1}{2}\;q_{12}^{\mu} q_{23}^{\nu}(q_{12}+ q_{23})^{\lambda}\;
\langle \,(x-\langle x \rangle)_{\mu}\;
(x-\langle x \rangle)_{\nu}\; (x-\langle x
\rangle)_{\lambda}\, \rangle
+{\cal O}(q_{ij}^4)\;.
\label{bigphi}
\end{eqnarray}
This result was also obtained by Heinz and Zhang~\cite{HZ}. 
To obtain this formula we made use of the triangle relation (\ref{triangle})
which assures that the three linear terms sum up to zero. The
on mass shell constraint in (\ref{massshell}) fixes the
time components of the four vectors. These redundant components 
are not explicitly eliminated.

The last term in Eq. (\ref{bigphi}) is of order $(qR)^3$ and is
generic. In contrast
the first two terms are a result of choosing the three momenta $K,
q_{12}$ and $q_{23}$ to span our 9 dimensional coordinate space and are
of order $q^2R/K$ and $q^3 R/K^2$ respectively. Since typical particle 
momenta are $k_i\simeq K\sim 300$ MeV and typical heavy
ion sources in nuclear collisions have a size $R\sim5$ fm,
we find that $q/K \ll qR \sim \hbar$, so that the generic
$(qR)^3$-contribution can actually dominate in the experimentally
relevant momentum regime.

In a simple but realistic model for flow, we will demonstrate in
the next section why contributions of order $q^2R/K$ are so
small. Afterwards we check in two further models for resonance and
source geometry if we actually can produce a significant phase effect
at scales of order $qR \sim \hbar$.

%%%%%%%%%%%%%%%%%%%%%%%%%%%%%%%%%%%%%%%%%%%%%%%%%%%%%%%%%%%%%%%%%%%%%%%%%%%%%%%

\section{Asymmetric Sources}

 From Eq. (\ref{expphi}) and (\ref{bigphi}) we see that there are
contributions to the phase due to odd space-time moments of the
source emission function of order $q^2R/K$ and $(qR)^3$.
To evaluate these contributions we need a model for the source.  In
the following we will study a number physical effects, which are known
to be present in relativistic heavy ion collisions, and which
result in asymmetric sources.  The examples we will study in the next
four subsections are flow,
resonances, moving and bursting sources respectively. We will estimate
their quantitative influence on the phases $\phi_{ij}$.

\subsection{Flow}

The most common ansatz for particle production and
collision dynamics in ultrarelativistic heavy ion collisions is the
Bjorken scenario. One assumes
cylindrical symmetry and requires local thermal equilibrium
with longitudinal Bjorken flow ($u_z={\rm z}/t$) as well as transverse
flow ${\bf v}$ through a Boltzmann factor. Thus
\begin{equation}
   S(x,{\bf K}) \sim e^{-K\cdot u/T}\, S_x(x)\, . \label{S}
\end{equation}
where $u=\gamma(v)({\bf v},\sinh(\eta),\cosh(\eta))$ is
the flow four-vector so that
\begin{eqnarray}
K\cdot u&=&m_\perp\gamma(v)(\cosh(\eta-Y)-{\bf {\beta}}_K\cdot{\bf v})\,.
\label{Ku}
\end{eqnarray}
Here, $\tau=\sqrt{t^2-{\rm z}^2}$ is the invariant
time, $\eta= 0.5\ln(t+{\rm z})/(t-{\rm z})$ the space-time rapidity.
and ${\bf v}=(v_x,v_y)$ the transverse flow. The pair velocity
is ${\bf {\beta}_K} = {\bf K}/K^0$. 
The transverse flow contribution in
Eqs. (\ref{S}) and (\ref{Ku}) provide a strong
$K$-dependence. Further $K$-dependences in the
emitting source, $S_x(x)$, would introduce additional $K$-dependences.
All these $K$-dependences should contribute to the second order
kinematic correction in Eq. (\ref{bigphi}).

When transverse flow is present only a few terms contribute.
They all vanish though, if we boost
into the so called longitudinal center of momentum
frame (LCMS).
In this frame we have $Y=0$ and the center of momentum of the pair is
parallel to the $x$- direction. This direction is labeled the outward
direction (o), while the beam axis or $z$-direction is labeled the longitudinal
direction (l). Perpendicular to these is the sidewards (s) or $y$-direction.  
With this in mind we can evaluate the $(q/K)^2$ contribution of
Eq. (\ref{bigphi})
\begin{eqnarray}
q_{12}^{\mu} q_{23}^{\nu}\,
\left[\frac{\partial \langle x_{\mu} \rangle}{\partial K^{\nu}} 
-\frac{\partial \langle x_{\nu} \rangle}{\partial K^{\mu}} \right]
&=&-\frac{1}{T}\,q_{12}^{\mu} q_{23}^{\nu}\,
\left[ \langle x_{\mu}u_{\nu} \rangle -
\langle x_{\nu}u_{\mu} \rangle \right] = 0\;.
\label{order2}
\end{eqnarray}
In the last step we made use of the fact, that $\langle y \rangle =0$
due to cylindrical symmetry and reflection symmetry in the $x-z$ plane.
We also boosted to the LCMS in which $Y=0$, so that
$\langle z\rangle=\langle u_z\rangle=0$ and ${\bf v}=(v_x,0)$ points into
the outward direction. Furthermore, due to
symmetry any choice
for $v_x$ has to be an even function in $y$ and $z$, so that 
$\langle v_x z\rangle=\langle v_x y\rangle=0$. Finally, in the LCMS
the pair velocity simplifies, so that $q_{ij}^t=\beta_x q_{ij}^x$. As a
result the $xt-$ and $tx-$ contribution in Eq. (\ref{order2}) cancel
each other.

This result demonstrates, that the standard $K$-dependent contribution
to the source emission function, i.e., flow does not produce sizable
contributions to the kinematic correction of second order. This
correction actually vanishes in the LCMS. We will see in the following
examples that higher order, but generic, contributions like resonance
and source geometry can in principal dominate the experimentally
accessible momentum range.

\subsection{Resonances}

We can study the influence of resonances on the phase within a simple
model~\cite{resmodel}. 
Under the assumption of classical propagation, the resonance travels 
an extra distance $\Delta x=u_r \Delta
\tau$ before it decays and produces a pion. $\Delta \tau$ is hereby
the life time of the resonance and $u_r$ its velocity. If,
furthermore, the
resonance life time is exponentially distributed with a decay width
$\Gamma_r = 1/\tau_r$ we obtain after averaging over the resonance
life time a modified source emission function. Its Fourier transform
reads
\begin{eqnarray}
F_r({\bf q}_{ij}, {\bf K}_{ij}) \sim
\int d^4 x\;S(x,E_K,{\bf K}_{ij})\;\exp(i q_{ij} x)\;(1-i
		q_{ij} \cdot u_r \tau_r)^{-1}\;.
                         \label{modfs}
\end{eqnarray}
Every resonance $r$ will supply such a contribution to the correlation
function and $x$ refers
to the space--time production point of the resonance.

We can evaluate the phase in (\ref{phi}) for a source consisting of
one resonance only. For a resonance velocity
independent of the space--time production point $x$ we find
\begin{equation}
\tan{\phi_{ij}}=u_r \cdot q_{ij}\; \tau_r.
\label{resphi}
\end{equation}
We would like to investigate the significance of this result 
and compare it with current experiments. Due to the limited statistics
in the experiments one reduces the 9 dimensional momentum
space, on which the cosine defined in Eq. (\ref{c3p}) depends,
down to one invariant momentum
\begin{equation}
 Q_{\rm 3}^2 = q_{\rm 12}^2 + q_{\rm 31}^2 + q_{\rm 23}^2\;,
   \label{q3}
\end{equation}
which is used to analyse the data.

In a previous paper we evaluated the radial part of the Fourier
transform $F$, neglecting any phase contribution and found that the 
correlation function in
Eq. (\ref{c3p}) can be rewritten as
\begin{eqnarray}
C_{\rm 3}(Q_{\rm 3}) = 1
  &+& 3\;\lambda_2\;\exp(-\frac{x^2}{3})\;(1+{\cal O}(x^4)) \nonumber \\
  &+& 2\;\lambda_2^{1.5}\exp(-\frac{x^2}{2})\;(1+{\cal O}(x^4))\;
\cos{(\phi_{12}+\phi_{23}+\phi_{31})}\;,
\label{c3approx}
\end{eqnarray}
where $x= Q_{\rm 3}\;R_{av}$ and $R_{av}$ is an average source emission
size parameter. $R_{av}$ is obtained by forming the average mean
square of the experimentally measured radii from two particle
correlations. The parameter $\lambda_2$ accounts
phenomenologically for  a number of effects like a
partially coherent sources, long lived resonances, final state
interactions and Coulomb screening effects. These effects tend to reduce the
value of this parameter form its ideal value $1$. It's value is also taken
from the experimentally determined two particle correlation functions.
The corrections of order $x^4$ are due to the
non--spheriocity of the source emission function. These corrections are
found to be rather small experimentally, in the order of a
percent, {\it i.e.} the source is close to spherical.

In calculating the phase we consider for simplicity 
only the momentum component in detector
or ${\bf K}_{ij}$-direction, i.e. the outward component $q_{ij,o}$, and
set all other momentum contributions to zero. With this
simplifications we can evaluate the cosine in terms of only the
outwards momenta
\begin{eqnarray}
\lefteqn{\langle\cos{(\phi_{12}+\phi_{23}+\phi_{31})\rangle}
(Q_{\rm 3,o}) = } \nonumber\\
& & \frac{
\int dq_{\rm 12,o}\;dq_{\rm 31,o}\;dq_{\rm 23,o}\; 
\cos{(\phi_{12}+\phi_{23}+\phi_{31})}
\;\delta(Q_{\rm 3,o}^2-q_{\rm 12,o}^2-q_{\rm 31,o}^2-q_{\rm 23,o}^2 )
\;\delta (q_{\rm 12,o}+q_{\rm 31,o}+q_{\rm 23,o})}
{\int dq_{\rm 12,o}\;dq_{\rm 31,o}\;dq_{\rm 23,o}\; 
\delta(Q_{\rm 3,o}^2-q_{\rm 12,o}^2-q_{\rm 31,o}^2-q_{\rm 23,o}^2)\;
\delta (q_{\rm 12,o}+q_{\rm 31,o}+q_{\rm 23,o})}.
\label{rescos}
\end{eqnarray}
The $\delta$--function of the sum of the relative
outward momenta assures that the 3 particles span a triangle.

In figure 1 we plot the cosine defined in Eq. (\ref{rescos}) for
different resonances. At large relative momentum each phase approach
$\pi/2$ as seen from Eq. (\ref{resphi}).
Therefore the cosine of the three phases vanish for
large $Q_{\rm 3,o}$ on a scale set by the
decay width $1/\tau_r$. 
In the momentum region where
correlation functions are sizeable,
$Q_{ij} \la 50$ MeV, the short lived resonances as the
$K^{\star}, \Delta$ and $\rho$ do not produce a sizeable phase.
The long lived resonances as $\eta$,
$\eta'$ and $K_0^S$ cannot be seen at all due to their
small form factor $\sim (1+u_r\cdot q\tau_r)^{-2}$ which effectively
removes these long lived resonances from Bose-Einstein correlations
and leads to an effectively smaller $\lambda$ \cite{resmodel}.
Only the $\omega$ could have a
significant influence on the phase at momenta $Q_{ij} \sim 50$ MeV.
However, it would be small due to its own form factor and due to the
small fraction of pions that are decay products of  $\omega$'s.
\begin{figure}
\centerline{
\psfig{figure=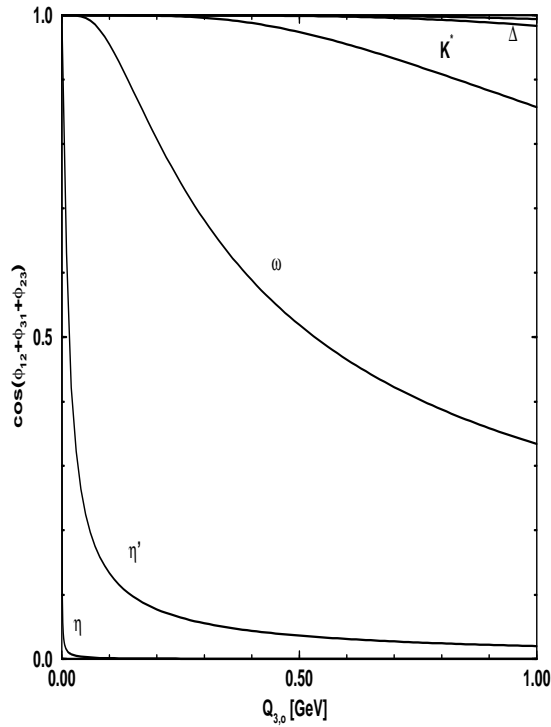,height=8.5cm,width=11.5cm,angle=-90}}
\caption{The phase contribution for different
resonances as a function of the momentum 
$Q_{\rm 3,o}^2=q_{\rm 12,o}^2+q_{\rm 13,o}^2+q_{\rm 23,o}^2$. The
fall-off of the curves scales with the resonance life time. The
transverse momentum of the pion pair is chosen $\beta_{\perp}=0.7$.
}
\end{figure}
 
The important part to notice is, that the phase contribution
is suppressed additionally by a Gaussian form factor as in
Eq. (\ref{c3approx}). Even if the
cosine is sizeable, like for example for the $\omega$, the overall
modification to the measured correlation function is still rather small
due to the form factor suppression. Only an experiment with
excellent resolution will be able to resolve the resonance contribution.

\subsection{Moving surfaces}

To have a detectable contribution to the phase we need 
to create an odd modification of the source emission
function, which appears at a momentum scale small enough, so
that it is not suppressed to drastically by the Gaussian form factor.
On the other hand such a contribution should be weighted 
strong enough to overcome the suppression for
small momenta shown in Eq. (\ref{expphi}). 
In this section we try to construct an odd source geometry, which has
this properties. 

Inspired by hydrodynamical models we assume a source which predominantly emits
particles from a thin surface layer. Such a source emission
function can be described as
\begin{equation}
S_A(x) \sim \delta(R(\tau)-r_\perp)\;,
\end{equation}
where $r_\perp$ is the transverse radius and $\tau$ is the proper
time. The source is cylindrically
symmetric with a radius $R(\tau)=R_0\;(1-(\tau/\tau_f)^\alpha)$. The
transverse radius starts out at $R_0$ and the source disintegrates
at time $\tau_f$.

It is straightforward to evaluate the phase such an extreme geometry
produces. We find
\begin{eqnarray}
\tan(\phi_\alpha) = \frac{a\;\cos a-\sin a +\sum_{n=0}^{\infty}\;
\frac{(-1)^n\; a^{2n+3}}{(2n+1)!\;(2n+\alpha+3)}}
{\cos a+a\;\sin a -1-\sum_{n=0}^{\infty}\;
\frac{(-1)^n\; a^{2n+2}}{(2n)!\;(2n+\alpha+2)}}\;,
\label{geophi}
\end{eqnarray}
if we assume, that the outward component of the momentum is
dominant, i.e. if we neglect the other two components. 
The only remaining variable is $a=\beta_{\perp} q_{ij,o}
\tau_f$,
where $\beta_{\perp}$ is the transverse velocity of the pion pair.
For $\alpha=1$ we find $\phi_1=-a/2$ and in the limit of $a \rightarrow
0$ we obtain 
\begin{equation}
\phi_{\alpha}= -\frac{2}{3}\;a\;\frac{2+\alpha}{3+\alpha} + {\cal O}(a^3)\;.
\label{asylim}
\end{equation}
The term linear in $a$ vanishes once the cosine of the sum of the three
phases is evaluated due to the triangle constraint, Eq. (\ref{triangle}). 
We plot in figure 2 the difference $\Delta \phi$ 
between the phase given in Eq. (\ref{geophi}) and 
the linear term of Eq. (\ref{asylim}). This difference is a direct measure for
the strength of the cubic correction in Eq. (\ref{expphi}). 
\begin{figure}
\centerline{
\psfig{figure=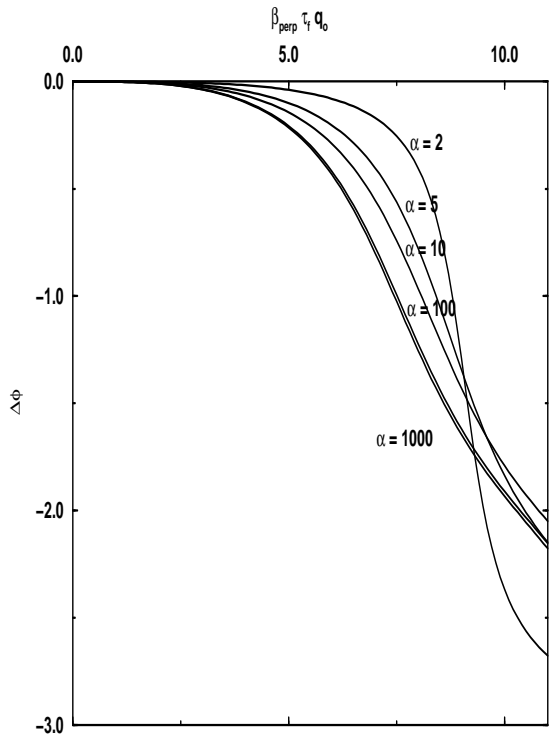,height=8.5cm,width=11.5cm,angle=-90}}
\caption{The difference $\Delta \phi$ 
between the phase $\phi_\alpha$ and its limit for small $q_o$ in
the case of surface emission.
}
\end{figure}

The contribution for $\alpha=1$ to the phase difference
vanishes identically. A surface, moving with constant speed
does not produce any phase. In this case the product $\tau R(\tau) \propto
\tau\; (\tau_f-\tau)$ is symmetric around $\tau_f/2$. The
additional factor of $\tau$ in the product comes from the integration
measure suitable for the Bjorken scenario. For accelerated surfaces, where
$\alpha\neq 1$ we find a sizable decrease at values $\beta_{\perp}
\tau_f q_0 \gg 5$. For typical values of the transverse momentum 
$\beta_{\perp} \sim 0.7$ and typical time scales $\tau_f$ of a few
Fermi this corresponds to outward momenta in the GeV region. We have
to rule out such a surface process.

\subsection{Bursts}

Another possibility to geometrically obtain odd modification functions
to the source emission is asymmetric bursts
\begin{eqnarray}
S_{V,2}(x) \sim \frac{1}{2}\;\left[(1-\epsilon)\;\delta(t-t_c)+
(1+\epsilon)\;\delta(t-t_c-\;\Delta t)\right]\; S_s({\bf r}).
\end{eqnarray}
This source corresponds to two bursts of particles emitted at times
$t_c$ and $t_c+\;\Delta t$ from a spatial emission
source function, $S_s({\bf r})$; the latter is arbitrary and irrelevant 
as long as it is symmetric. 
The two contributions are weighted by $\epsilon$ such
that $|\epsilon|\le1$. 

The phase for such a source is from Eq. (\ref{phi})
\begin{eqnarray}
\tan{\left(\phi-q_4 (t_c+\Delta t/2)\right)}=\epsilon\;\tan{(q_4 \Delta t/2)}
  \;.   \label{phisum}
\end{eqnarray}
where $q_4=E_i-E_j\simeq \beta_\perp q_o$.
The contribution proportional to the mean time $q_4\;(t_c+\Delta t/2)$
on the right hand side cancels when the three phases are added
due to the triangle
constraint (\ref{triangle}). Thus the phase effectively vanishes for equal
weights $\epsilon=0$, if one weight vanishes $\epsilon=\pm 1$ or for
zero time separation of the two flashes. This is simply because the
source is symmetric in all these cases.
The scale of variation of the phase is given
by the $\Delta t$ and the size of the phase is given by the asymmetry
parameter $\epsilon$. 
\begin{figure}
\centerline{
\psfig{figure=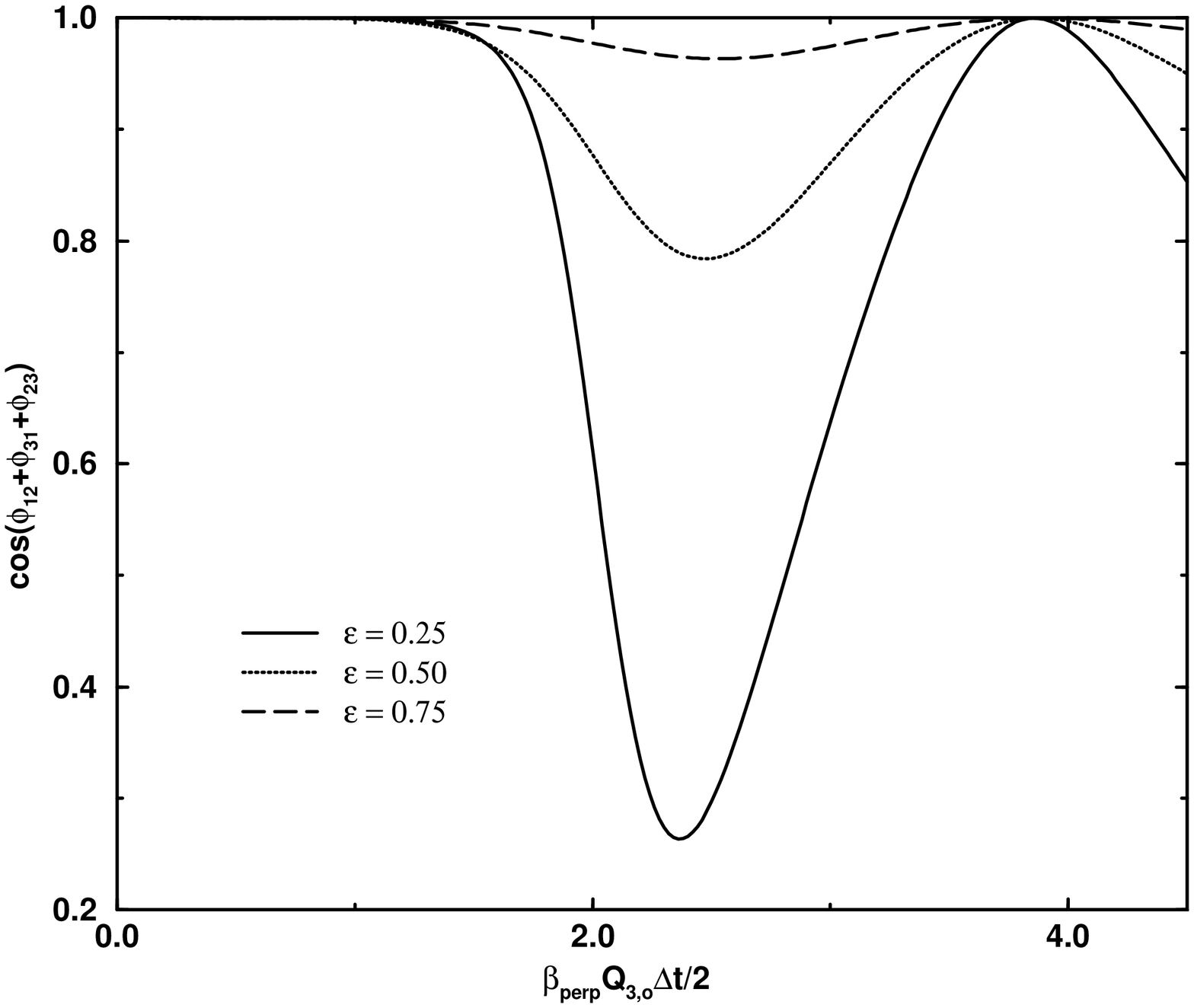,height=8.5cm,width=11.5cm,angle=-90}}
\caption{
The phase contribution in the case of volume
emission from two instantaneous sources separated by a time interval
$\Delta t$. The weight $\epsilon$ is set to $.25, .5$ and $.75$.
}
\end{figure}

In figure 3 we plot the cosine of the three phases according to
Eq. (\ref{rescos}) for different values of $\epsilon$, assuming again
that the outward component of the relative momentum is dominant. The
phase scales in the variable $\beta_\perp Q_{3,o} \Delta t$. 
For outward momenta close to $\beta_\perp Q_{3,o} \Delta t/2\sim2$ one of
the phases in Eq. (\ref{phisum}) 
changes sign and we see a strong signal. In the
experimentally accessible momentum range this would correspond to a
temporal separation of the bursts in the order of 10 fm/c. While such
a scenario is rather unrealistic it might open in the long run some
new ideas and approaches to investigating possible phase signals.

\section{Conclusion}

We have studied 3-body correlations for incoherent sources which can be
calculated from 2-body correlations except for a phase. 
The phase is due to odd space-time moments of the
source emission function and vanish at small momentum transfer to
order $(qR)^3$
and $q^2R/K$, where the former dominates in the experimentally
relevant momentum regime in relativistic heavy ion collisions.

Effects of flow, resonances,
moving and bursting sources were studied as they are known to be present in
relativistic heavy
ion collisions, and result in asymmetric sources.
Their influence on the phases $\phi_{ij}$ was only significant at large
relative momenta where the form factors, $F(q)$, and thus also correlation
functions were small or in some extreme scenarios, where all particles
where emitted in 2 sudden bursts. 
In both cases, the experimental extraction of the phases
needs very high resolutions. The phase in 3-body correlation functions
seems to be very elusive.

\section*{Acknowledgements}

We would like to thank Ulrich Heinz for helpful discussions
and Janus Schmidt--S{\o}rensen for providing
and explaining the necessary data. 
\\

%\newpage
%%%%%%%%%%%%%%%%%%%%%%%%%%%%%%%%%%%%%%%%%%%%%%%%%%%%%%%%%%%%%%%%%%%%%%%%%%%%%%

\section*{Figure Captions}

\noindent {\bf Figure 1.} The phase contribution for different
resonances as a function of the momentum 
$Q_{\rm 3,o}^2=q_{\rm 12,o}^2+q_{\rm 13,o}^2+q_{\rm 23,o}^2$. The
fall-off of the curves scales with the resonance life time.
The transverse momentum of the pion pair is chosen $\beta_{\perp}=0.7$.\\ 

\noindent {\bf Figure 2.} The difference $\Delta \phi$ 
between the phase $\phi_\alpha$ and its limit for small $q_o$ in
the case of surface emission.\\

\noindent {\bf Figure 3.} The phase contribution in the case of volume
emission from two instantaneous sources separated by a time interval
$\Delta$. The weight $\epsilon$ is set to $.25, .5$ and $.75$.\\
\\
%%%%%%%%%%%%%%%%%%%%%%%%%%%%%%%%%%%%%%%%%%%%%%%%%%%%%%%%%%%%%%%%%%%%%%%%%%%%%%%%%%%%%%%%%%%%%%%%%%%%%
%  ---------------------------------------------------------------------
%  REFERENCES
%  ---------------------------------------------------------------------
%\newpage


\baselineskip18pt
\begin{thebibliography}{99}
\bibliographystyle{unsrt}

\bibitem{janus} H. Becker {\it et al.}, NA44, {\em to be published},
                Janus Schmidt--S\"orensen, thesis, Niels Bohr
                Institute (1995)

\bibitem{hv1} H. Heiselberg and A. P. Vischer, {\em Phys. Rev.} {\bf
C55} 874, (1997).

\bibitem{HZ} U. Heinz and Q.H. Zhang, nucl-th/9701023

\bibitem{cramer} J. G. Cramer, {\em Phys. Rev.} {\bf C43}, 2798,
(1991); J. G. Cramer and K. Kadija, {\em Phys. Rev.} {\bf C53}, 908 (1996) .

\bibitem{biya} M. Biyajima, A. Bartl, T. Mizoguchi, O. Terazawa and N. Suzuki,
{\em Prog. Theor. Phys.} {\bf 84}, 931 (1990); N. Suzuki and
M. Biyajima, {\em Prog. Theor. Phys.} {\bf 88}, 609 (1992).

\bibitem{weiner} R.M. Weiner, {\em Phys. Lett.} {\bf B232}, 278
(1989); {\bf B242}, 547 (1990); M. Pl{\"u}mer, L.V. Razumov and   
R.M. Weiner, {\em Phys. Lett.} {\bf B286}, 335 (1992); I.V. Andreev, 
M. Pl{\"u}mer, and R.M. Weiner, {\em Int. J. Mod. Phys.} {\bf A8}, 4577
(1993).

\bibitem{def} M. Gyulassy, S. Kaufmann and L. Wilson, {\em Phys. Rev.}
{bf C20}, 2269, (1979).

\bibitem{Heinz} S. Chapman, J.R. Nix, and U. Heinz,
{\em Phys. Rev.} {\bf C52}, 2694 (1995); S. Chapman, U. Heinz and P. Scotto,
{\em Heavy Ion Physics} {\bf 1},1 (1995).

\bibitem{resmodel} M. Gyulassy and S. S. Padula, {\em Phys. Rev.} {\bf
C41}, R21 (1990); H. Heiselberg, {\em Phys. Lett.} {\bf B379}, 27 (1996).  


\end{thebibliography}
\end{document}